\documentclass[11pt]{article}
\usepackage[margin=1in]{geometry}
\usepackage{amsmath,amssymb,amsthm,mathtools}
\usepackage[shortlabels]{enumitem}
\usepackage[expansion=false]{microtype}
\usepackage{graphicx}
\usepackage{booktabs}
\usepackage{xcolor}
\usepackage{tikz}
\usetikzlibrary{arrows.meta,calc,decorations.markings}
\usepackage[font=small,labelfont=bf]{caption}
\usepackage[colorlinks=true,linkcolor=blue!50!black,citecolor=green!40!black,urlcolor=blue!50!black]{hyperref}

\numberwithin{equation}{section}
\newtheorem{theorem}{Theorem}[section]
\newtheorem{proposition}[theorem]{Proposition}

\theoremstyle{remark}
\newtheorem{remark}[theorem]{Remark}

\newcommand{\Hil}{\mathcal H}
\newcommand{\Hphy}{\mathcal H_{\rm phy}}
\newcommand{\Ps}{\mathcal P}
\newcommand{\Pphy}{\mathcal P_{\rm phy}}
\newcommand{\Padm}{\mathcal P_{\rm adm}}
\newcommand{\Cq}{\mathcal C}
\newcommand{\K}{\mathcal K}
\newcommand{\N}{\mathcal N}
\newcommand{\D}{\mathcal D}
\newcommand{\R}{\mathbb R}
\newcommand{\C}{\mathbb C}
\newcommand{\Z}{\mathbb Z}
\newcommand{\CP}{\mathbb{C}P}
\newcommand{\PP}{\mathbb P}
\newcommand{\Chat}{\widehat C}
\newcommand{\Phat}{\widehat P}
\newcommand{\Ohat}{\widehat O}
\newcommand{\Fhat}{\widehat F}
\newcommand{\eps}{\varepsilon}
\newcommand{\ip}[2]{\langle #1|#2\rangle}
\newcommand{\ev}[1]{\langle #1\rangle}
\newcommand{\vthree}{\cite{Corichi2008}}
\DeclareMathOperator{\Tot}{Tot}
\DeclareMathOperator{\Hom}{Hom}
\DeclareMathOperator{\spec}{spec}
\DeclareMathOperator{\Dom}{Dom}

\title{Group averaging and the geometry of physical quantum states:\\
from normalizable solutions to solvable loop quantum cosmology}
\author{Alejandro Corichi\\[4pt]
\small Centro de Ciencias Matem\'aticas, Universidad Nacional Aut\'onoma de M\'exico,\\
\small UNAM-Campus Morelia, A.~Postal 61-3, Morelia, Michoac\'an 58190, Mexico\\
\small \texttt{corichi@matmor.unam.mx}}
\date{}

\begin{document}
\maketitle

\begin{abstract}
In canonical quantum gravity the physical states solve the constraints but are typically not normalizable in the kinematical Hilbert space. Group averaging constructs them from normalizable kinematical seeds. We describe the geometry of this seed-to-state map in the language of geometric quantum mechanics, complementing the description of physical states as the zero set of the squared constraint.

When averaging is an orthogonal projection, the admissible kinematical rays form a holomorphic vector bundle over the physical ray space, whose zero section consists of the embedded exact solutions. Its fibres contain the gauge orbits, and their complexifications where these are defined, but are in general strictly larger. The kinematical symplectic form does not descend, and a unitary flow descends to the base exactly when its generator commutes with the averaging projection, equivalently when the Hessian of the squared constraint vanishes along it.

For distributional physical states, group averaging on an explicit Sobolev test space yields an affine bundle of representatives with no zero section. Sections given by a fixed seed profile are K\"ahler-isometric embeddings of the physical ray space whose images contain no exact solutions. The test-space inner product singles out one of them, and the gradient flow of the squared constraint moves seeds along the fibres, producing Weyl sequences of approximate solutions.

We apply this to solvable loop quantum cosmology. Group averaging reproduces the Klein--Gordon physical inner product on an explicit test space. Because the scalar-field momentum enters the constraint quadratically, the constraint spectrum at each gravitational mode has an edge at a finite distance below the shell. With a test norm of fixed scale, the canonical section is then mode-dependent and not K\"ahler isometric, and the canonical seeds of modes near the edge are biased off the shell. No mode-wise reweighting of the test norm makes the canonical section both isometric and unbiased. The same quadratic dependence introduces the Jacobian that separates the Page--Wootters state from the slice of the Klein--Gordon solution. An ideal clock shows none of these effects.

The results separate the canonical structures, namely physical rays, physical geometry and Dirac observables, from those that depend on a choice of seeds, test space or clock.
\end{abstract}

\noindent\textbf{Keywords:} group averaging, refined algebraic quantization, loop quantum cosmology, geometric quantum mechanics, quantum constraints, Dirac observables, problem of time

\tableofcontents

\section{Introduction}\label{sec:intro}

Most of the fundamental theories of physics, when written in Hamiltonian form, are subject to constraints among their canonical variables. Gauge theories are the prototypical example, and general relativity is the most important one: its Hamiltonian is itself a combination of constraints, and the dynamics it describes is, in a precise sense, pure gauge. In the Dirac approach to the quantization of such systems one first constructs a kinematical Hilbert space, on which the constraints are represented as operators, and then looks for the physical states, those annihilated by the constraints \cite{Dirac,HT}. The space of solutions must then be endowed with a new, physical, inner product.

It is well known that this second step is far from trivial. In canonical quantum gravity the Hamiltonian constraint generates neither a compact group nor, in general, a group action with normalizable invariant vectors. The physical states are then not elements of the kinematical Hilbert space, and one needs a procedure to construct them. Group averaging, the central tool of refined algebraic quantization \cite{Higuchi,ALMMT,GiuliniMarolf1,GiuliniMarolf2,Marolf}, does precisely that. Starting from a normalizable kinematical state, which we shall call a \emph{seed}, one averages it over the orbit of the constraints, obtaining a solution, possibly distributional, together with an inner product on the space of solutions. The same construction underlies the comparison of physical inner products and of the different descriptions of evolution in loop quantum cosmology \cite{KLP} and in relational quantum dynamics \cite{HSL1,HSL2}.

There is a natural arena in which to study these constructions, namely the space of quantum states itself. In the geometric formulation of quantum mechanics \cite{Kibble,CMP,AshtekarSchilling,BrodyHughston} the space of rays of the Hilbert space is a K\"ahler manifold. Its symplectic structure implements unitary transformations as canonical transformations, while its Riemannian metric encodes uncertainties and transition probabilities. One can then ask, just as one does in the classical theory, what the constraints do to the geometry.

This question was raised in \cite{Corichi2008}. There it was observed that, when the physical states are normalizable, the Dirac conditions can be written, as in the classical theory, as the vanishing of a function on the space of rays: the expectation value of the square of the constraint. The physical state space is then a submanifold of the kinematical one. In contrast with the classical case, however, the induced symplectic form has no degenerate directions, and no further quotient is needed. It was also stressed there that group averaging should not be thought of as a quotient by `quantum gauge orbits': averaging adds vectors, and it is a projection rather than an identification. The revised version of \cite{Corichi2008} turns these observations into theorems. Among other things, it shows that the expectation value of the squared constraint is a Morse--Bott function whose transverse Hessian is the spectrum of the squared constraint operator; that the double-commutator condition on observables is equivalent to their commuting with the projector onto physical states; that the gradient flow of this function is heat-kernel smeared group averaging and converges to the nearest physical ray, along fibres that are invariant under the complexified gauge group; and that, when zero lies in the continuous spectrum, the same flow produces Dirac states and, through its decay rate, the rigging map of refined algebraic quantization.

In this article we look at the same structures from a different angle. Our object of study is the map that takes a seed to the physical state it produces. Group averaging assigns the same physical ray to many seeds. What is the resulting equivalence class? How is it related to the embedded physical space? What remains of this structure when no physical state is normalizable? And what changes when the clock momentum enters the constraint quadratically, as it does in cosmology? These are the questions we want to address. As we shall see, the answers separate quite neatly the structures that are canonical, namely the physical rays, their K\"ahler geometry and the Dirac observables, from those that depend on additional choices: a family of seeds, a test space, or a clock.

Let us summarize our main results.
\begin{enumerate}[(i)]
\item When averaging is an orthogonal projection, the admissible kinematical ray space is a holomorphic vector bundle over the physical ray space, with the embedded physical space as its zero section (Proposition~\ref{prop:bundle}; the underlying linear projection appears in \vthree). Its fibres contain the gauge orbits and their complexifications, but are in general strictly larger (Section~\ref{sec:orbits}). The ambient symplectic form does not descend, while the physical one pulls back with kernel exactly the fibres (Proposition~\ref{prop:forms}).
\item A unitary flow descends to the bundle base exactly when its generator commutes with the averaging projection (Proposition~\ref{prop:descent}). For bounded generators this is the double-commutator criterion of \vthree, which Proposition~\ref{prop:hessian} recalls as the vanishing of the Hessian of the squared constraint along the flow, and extends to unbounded constraints.
\item In the fixed-energy two-oscillator model of \cite{ABC}, the full bundle of seeds and the bundle of coherent seeds differ. Coherent seeds average to spin coherent states, which form a one-dimensional subvariety of the $N$-dimensional physical space, and within the coherent family each averaging fibre is a single complexified gauge orbit (Section~\ref{sec:oscillators}).
\item For a parametrized oscillator, whose physical states are distributional, averaging on an explicit Sobolev test space yields an affine bundle of representatives (Proposition~\ref{prop:affine}). Sections defined by a fixed seed profile are K\"ahler-isometric embeddings of the physical ray space whose images contain no exact solutions. The test-space inner product selects one of them, and the gradient flow of the squared constraint moves seeds along the fibres, producing Weyl sequences of approximate solutions (Proposition~\ref{prop:vertical}). These lifts are distinct from clock-conditioned states.
\item In solvable loop quantum cosmology (Section~\ref{sec:slqc}), group averaging on an explicit test space reproduces the Klein--Gordon physical inner product (Proposition~\ref{prop:slqc-raq}). Because the clock momentum enters quadratically, the constraint spectrum has an edge at distance $\kappa_k=k^2/\eps^2$ below the shell. This edge makes the canonical section mode-dependent, prevents it from being K\"ahler isometric, and biases the canonical seeds of modes near the edge off the shell (Proposition~\ref{prop:edge}); no mode-wise reweighting of the test norm removes both effects (Proposition~\ref{prop:weights}). The same quadratic dependence introduces the Jacobian $2\sqrt\Theta$ that separates the Page--Wootters state from the slice of the solution.
\end{enumerate}

Related approaches impose the constraints through spectral projections of the constraints themselves, as in Klauder's projection-operator method \cite{Klauder1,Klauder2}, or through a single master constraint \cite{Thiemann}; the clock model makes the connection explicit. A complementary question is how the unreduced classical geometry, with its gauge directions, can be recovered from the quantum theory when the latter is regarded as fundamental. That question is addressed in \cite{CorichiGauge}, where the gauge directions are shown to be encoded in the linearization of the master-constraint flow and in the fluctuations of kinematical semiclassical states. The two articles can be read independently. We shall see in Section~\ref{sec:complexified} that they meet at a precise point: the complexified gauge directions that group averaging discards are the same directions that the master flow relates.

Throughout, $\Ps$ denotes the kinematical ray space, $\Pphy$ its physical counterpart, and $(g,\Omega)$ the quantum metric and symplectic form. The structure of the paper is as follows. In Section~\ref{sec:embedded} we recall the embedded description of the physical states and the Hessian criterion for observables. Section~\ref{sec:bundle} constructs the averaging bundle when averaging is an orthogonal projection, and Section~\ref{sec:oscillators} illustrates it with two oscillators at fixed total energy. Section~\ref{sec:clock} analyses the distributional case for a clock constraint with an ideal clock, and Section~\ref{sec:slqc} treats solvable loop quantum cosmology, where the clock is relativistic. We end with a discussion in Section~\ref{sec:discussion}.

\section{The embedded physical state space}\label{sec:embedded}

In this section we recall the geometric formulation of quantum mechanics and the description of normalizable physical states as the zero set of a function on the space of rays. We follow \cite{AshtekarSchilling} for the formalism and \vthree{} for its application to constraints.

\subsection{Quantum geometry}

Let $\Hil$ be a complex Hilbert space and $S\subset\Hil$ its unit sphere. The space of pure states is $\Ps=S/U(1)$, equivalently the projective space $\PP(\Hil)$ of nonzero vectors modulo nonzero complex factors. The complex structure $J$ on the underlying real Hilbert space is multiplication by $i$. Following \cite{AshtekarSchilling}, we decompose the inner product, antilinear in its first entry, as
\begin{equation}\label{eq:hermitian}
\ip{\Psi}{\Phi}=\frac{1}{2\hbar}\bigl(G(\Psi,\Phi)+i\,\Omega_\Hil(\Psi,\Phi)\bigr),\qquad \Omega_\Hil(X,Y)=G(JX,Y).
\end{equation}
Horizontal variations on $S$, those orthogonal to the phase direction, induce a metric $g$ and a symplectic form $\Omega$ on $\Ps$. For a nonzero holomorphic local representative $\Psi(z)$,
\begin{equation}\label{eq:kahlerpot}
\Omega=i\hbar\,\partial\bar\partial\log\|\Psi(z)\|^2 ,
\end{equation}
so that $g=2\hbar\,g_{\rm FS}$ and $\Omega=2\hbar\,\omega_{\rm FS}$, with $\omega_{\rm FS}=\frac i2\partial\bar\partial\log\|\Psi\|^2$ and $g_{\rm FS}$ the Fubini--Study K\"ahler form and metric in their standard normalization. The triple $(g,\Omega,J)$ makes $\Ps$ a K\"ahler manifold.

A self-adjoint operator $\Fhat$ defines the function $f([\Psi])=\ip{\Psi}{\Fhat\Psi}$ on normalized vectors in its domain. With the convention $\iota_{X_f}\Omega={\rm d}f$, its Hamiltonian vector field is represented horizontally by
\begin{equation}\label{eq:hvf}
X_f\big|_{[\Psi]}=-\frac1\hbar\,J(\Fhat-f)\Psi ,
\end{equation}
and generates the ray motion induced by $e^{-it\Fhat/\hbar}$. These are quantum Hamiltonian motions on $\Ps$, not motions on a classical constraint surface. The length of $X_f$ measures the uncertainty of $\Fhat$: $g(X_f,X_f)=(2/\hbar)(\Delta\Fhat)^2$, and $X_f$ vanishes exactly at the eigenrays of $\Fhat$.

\subsection{Constraints}

For a self-adjoint constraint $\Chat$, introduce the functions $c=\ev{\Chat}$ and
\begin{equation}\label{eq:Cq}
\Cq=\ip{\Psi}{\Chat^2\Psi}=\|\Chat\Psi\|^2 .
\end{equation}
The last expression is the closed quadratic form of $\Chat^2$, defined on $\Dom\Chat$ and set equal to $+\infty$ outside it. For normalized vectors, $\Cq=0$ is equivalent to $\Chat\Psi=0$, whereas $c=0$ is not sufficient. The difference between the two is the content of the decomposition \vthree
\begin{equation}\label{eq:decomp}
\Cq=c^2+\frac\hbar2\,g(X_c,X_c) ,
\end{equation}
which is nothing but $\ev{\Chat^2}=\ev{\Chat}^2+(\Delta\Chat)^2$ written geometrically. The first term mirrors the classical constraint. The second is the squared speed of the quantum gauge orbit through $[\Psi]$, and its vanishing forces that orbit to collapse to a point.

When the solution space $\Hphy=\ker\Chat$ is nonzero, it is closed, and we set $\Pphy=\PP(\Hphy)$. For several constraints $\Chat_I$ one uses $\Cq=\sum_I\|\Chat_I\Psi\|^2$, which is the expectation value of Thiemann's master constraint $\widehat M=\sum_I\Chat_I^2$ \cite{Thiemann}. The embedded physical space is thus the zero set of the master-constraint function on $\Ps$. When zero is an isolated eigenvalue, $\Cq$ is a Morse--Bott function whose Hessian in the directions normal to $\Pphy$ is given by the nonzero spectrum of $\widehat M$ \vthree. The inclusion $\iota:\Pphy\hookrightarrow\Ps$ is a K\"ahler isometry,
\begin{equation}\label{eq:isometry}
\iota^*g=g_{\rm phy},\qquad \iota^*\Omega=\Omega_{\rm phy}.
\end{equation}
In particular, $\Omega_{\rm phy}$ has no constraint-induced degeneracy, and on physical states the constraint field (\ref{eq:hvf}) vanishes. Let us note that invariance of a kinematical ray under the constraint motion is a weaker condition, since every eigenray of $\Chat$ is invariant, including those with nonzero eigenvalue. It is the condition $\Cq=0$ that selects the physical sector.

\subsection{Observables and the Hessian of the squared constraint}\label{sec:hessian}

Since $\Cq\ge0$ attains its minimum on $\Pphy$, its first derivative along any flow vanishes there, and it is the second derivative that decides whether a flow stays on the physical submanifold. The following proposition makes this precise.

\begin{proposition}\label{prop:hessian}
Let $u\in\Hphy$ be normalized and let $\Ohat$ be bounded and self-adjoint. Set $q_u(\Ohat)=\|\Chat\Ohat u\|^2\in[0,\infty]$, with the convention of (\ref{eq:Cq}).
\begin{enumerate}[(i)]
\item If $\Ohat u\in\Dom\Chat$, then, as a quadratic form,
\begin{equation}\label{eq:doublecomm}
\ip{u}{[\Ohat,[\Ohat,\Chat^2]]\,u}=-2\,q_u(\Ohat).
\end{equation}
\item Along $\Psi_t=e^{-it\Ohat/\hbar}u$ one has $\liminf_{t\to0}\Cq(\Psi_t)/t^2\ge q_u(\Ohat)/\hbar^2$. If $\Chat$ is bounded, $\Cq(\Psi_t)$ is smooth and $\frac{{\rm d}^2}{{\rm d}t^2}\Cq(\Psi_t)\big|_{t=0}=2q_u(\Ohat)/\hbar^2$.
\item $q_u(\Ohat)=0$ for all $u\in\Hphy$ if and only if $\Ohat\Hphy\subset\Hphy$, and then $\Hphy$ reduces $\Ohat$.
\end{enumerate}
\end{proposition}

\begin{proof}
(i) Expanding $[\Ohat,[\Ohat,\Chat^2]]=\Ohat^2\Chat^2-2\Ohat\Chat^2\Ohat+\Chat^2\Ohat^2$, the two outer terms vanish in the state $u$ because $\Chat u=0$, leaving $-2\ip{\Ohat u}{\Chat^2\Ohat u}$. (ii) Since $\Chat u=0$, $\Cq(\Psi_t)/t^2=\|\Chat(\Psi_t-u)/t\|^2$, and $(\Psi_t-u)/t\to-i\Ohat u/\hbar$ in norm. Lower semicontinuity of the closed form gives the inequality. For bounded $\Chat$ the expansion is smooth, and the second derivative is $-\hbar^{-2}\ip{u}{[\Ohat,[\Ohat,\Chat^2]]u}$. (iii) $q_u=0$ if and only if $\Ohat u\in\ker\Chat$. A bounded self-adjoint operator that preserves a closed subspace also preserves its orthogonal complement.
\end{proof}

Items (i) and (iii) are contained in \cite[Prop.~4.8]{Corichi2008}, where it is shown that, on physical states, the double-commutator condition of the master constraint programme \cite{Thiemann,DT1} is equivalent to $[\Ohat,\Phat]=0$, with $\Phat$ the projector onto $\Hphy$, and hence to the invariance of $\Pphy$ to all orders. Item (ii) extends the second-derivative statement to unbounded constraints. Geometrically, an observable preserves the physical submanifold exactly when the Hessian of $\Cq$ along its Hamiltonian flow vanishes on $\Pphy$, i.e.\ when $X_o$ is tangent to the zero set of $\Cq$. With several constraints the same statement holds for $\widehat M$, with $q_u=\|\widehat M^{1/2}\Ohat u\|^2$. In the distributional case of Section~\ref{sec:clock} there is no nonzero physical vector $u$ in $\Hil$, and the preservation of physical states must instead be expressed through the averaging map; see (\ref{eq:OA}) below.

\section{Orthogonal averaging as a vector bundle}\label{sec:bundle}

In this section averaging is the orthogonal projection $\Phat$ onto $\Hphy$. We first recall how such a projection arises, then describe the geometry of the map it induces on rays.

\subsection{Realizations of the orthogonal projection}

There are several standard situations in which group averaging is an orthogonal projection.

\paragraph{Compact groups.} If a compact group $G$ acts by a strongly continuous unitary representation $U$ whose invariant subspace is $\Hphy$, then, with normalized Haar measure,
\begin{equation}\label{eq:haar}
\Phat=\int_GU(s)\,{\rm d}\mu(s),\qquad \Phat^2=\Phat=\Phat^*.
\end{equation}
Invariance of the Haar measure makes every vector in the range invariant, averaging fixes every invariant vector, and invariance of the measure under inversion makes $\Phat$ self-adjoint. For a single constraint, $U(\theta)=e^{-i\theta\Chat}$, $\theta\in[0,2\pi)$, defines a $U(1)$ action only if $\spec\Chat\subset\Z$ in suitable units.

\paragraph{Mean ergodic averaging.} If $e^{-it\Chat/\hbar}$ defines an action of $\R$, the Ces\`aro averages
\begin{equation}\label{eq:cesaro}
A_T=\frac1{2T}\int_{-T}^Te^{-it\Chat/\hbar}\,{\rm d}t
\end{equation}
converge strongly, as $T\to\infty$, to the orthogonal projection onto $\ker\Chat$; this is von Neumann's mean ergodic theorem \cite{vonNeumann}. The analogous statement holds for F\o lner averages of unitary representations of amenable groups \cite{Greenleaf}. When $0$ is not an eigenvalue the limit is zero, and averaging must be redefined distributionally, as in Section~\ref{sec:clock}.

\paragraph{Heat-kernel averaging.} By the spectral theorem, $e^{-t\Chat^2}\to\Phat$ strongly as $t\to\infty$, for every self-adjoint $\Chat$. In \vthree{} it is shown that the corresponding rays are the gradient flow of $\Cq$ and that, for compact groups and for $\R^k$, $e^{-t\Chat^2}$ is the average of the constraint unitaries weighted with the heat kernel on the group. We return to this flow in Section~\ref{sec:vertical}.

\paragraph{L\"uders conditioning.} Because $\Phat$ is an orthogonal projection, $\Psi\mapsto\Phat\Psi/\|\Phat\Psi\|$ is the L\"uders state change \cite{Luders} conditioned on the outcome `constraint satisfied' of the measurement $\{\Phat,1-\Phat\}$. Its probability is the retained norm fraction $p$ of (\ref{eq:p}) below, which is also the squared overlap of the seed with its nearest physical ray \vthree.

\subsection{The bundle}

Write $\K=\ker\Phat$, so that $\Hil=\Hphy\oplus\K$. The map induced on rays is defined whenever the average is nonzero:\footnote{In the published version of \cite{Corichi2008} the symbol $\eta$ denotes the averaging projection; in its revised version and here it denotes the induced map on rays. It should not be confused with the rigging map of refined algebraic quantization \cite{ALMMT,GiuliniMarolf1}, also usually written $\eta$.}
\begin{equation}\label{eq:eta}
\eta:\Padm\longrightarrow\Pphy,\qquad \Padm=\Ps\setminus\PP(\K),\qquad \eta([\Psi])=[\Phat\Psi].
\end{equation}
The excluded rays are those annihilated by averaging. On $\Padm$, physical equivalence reads $[\Psi]\sim[\Phi]$ if and only if $\Phat\Psi=\alpha\,\Phat\Phi$ for some $\alpha\in\C^*$. Let us stress that this concerns the physical ray, not the equality of the kinematical states. Let $L\to\Pphy$ be the tautological line bundle, whose fibre at a physical ray $\ell$ is the line $\ell\subset\Hphy$, and let $\underline{\K}=\Pphy\times\K$ be the trivial bundle with fibre $\K$.

\begin{proposition}\label{prop:bundle}
The map (\ref{eq:eta}) is the projection of the holomorphic vector bundle
\begin{equation}\label{eq:bundle}
\Padm\;\simeq\;\Tot\Hom(L,\underline{\K})\;\cong\;\Tot(L^*\otimes\underline{\K}).
\end{equation}
Under this identification $\iota$ is the zero section and $\eta\circ\iota={\rm id}_{\Pphy}$.
\end{proposition}

\begin{proof}
Write $\Psi=u+k$, with $0\neq u=\Phat\Psi$ and $k\in\K$, and define $T:\ell=\C u\to\K$ by $T(u)=k$. Rescaling $\Psi$ rescales $u$ and $k$ together and leaves $T$ unchanged. Conversely, $(\ell,T)$ determines the ray $[u+T(u)]$ for any nonzero $u\in\ell$. For local triviality, choose a continuous linear functional $\varpi$ on $\Hphy$ and normalize physical representatives by $\varpi(u)=1$. The seed representatives are then $u+k$, with $k\in\K$ arbitrary; these are the standard projective charts of $\PP(\Hil)$ for the functional $\varpi\circ\Phat$. If a second normalization gives $u'=tu$, then $k'=tk$, which is the transition function of $\Hom(L,\underline\K)$, holomorphic and linear in the fibre. Finally, $T=0$ exactly when $[u]\in\Pphy$.
\end{proof}

The argument applies in finite and infinite dimensions. As a set, (\ref{eq:bundle}) is the familiar description of a linear projection between projective spaces \cite{GriffithsHarris}, and in this form it appears in \cite[Thm.~4.7]{Corichi2008}. What we want to emphasize is the vector-bundle structure itself, which has a distinguished zero section, the embedded exact solutions, and which will be lost in the distributional case. A fibre consists of all discarded components attached to a fixed physical line. A vector on that line identifies the fibre with $\K$, and changing the vector rescales the fibre coordinates.

\begin{remark}[Topology]
Like every vector bundle, (\ref{eq:bundle}) has a zero section; a trivialization would require a global frame. If $0<\dim\K=m<\infty$ and $\dim\Hphy\ge2$, the bundle is topologically nontrivial, since its restriction to any projective line $\CP^1\subset\Pphy$ is $\mathcal O(1)^{\oplus m}$, with first Chern class $m\neq0$ \cite{GriffithsHarris,MilnorStasheff}. If $\K$ is infinite-dimensional and separable, the structure group acts through the general linear group of $\K$, which is contractible by Kuiper's theorem \cite{Kuiper}, so the bundle is topologically trivial. Its holomorphic triviality is a separate question that we shall not address.
\end{remark}

\begin{figure}[t]
\centering
\begin{tikzpicture}[scale=0.95,>=Stealth]
\node[font=\small] at (2.6,4.35) {\textbf{(a)} orthogonal averaging};
\foreach \x in {0.5,1.5,2.5,3.5,4.5}{\draw[gray!60] (\x,0) -- (\x,3.3);}
\draw[very thick,red!70!black] (0,0) -- (5.0,0) node[right,font=\footnotesize,align=left]{$\iota(\Pphy)$\\[-1pt]zero section};
\foreach \x in {0.5,1.5,2.5,3.5,4.5}{\fill[red!70!black] (\x,0) circle (1.6pt);}
\foreach \x in {0.5,1.5,2.5}{\draw[->,thick,blue!60!black] (\x+0.12,2.9) -- (\x+0.12,0.4);}
\node[font=\footnotesize,blue!60!black] at (1.5,3.7) {flow of $\Cq$};
\node[font=\footnotesize,gray!30!black,anchor=south] at (3.5,3.3) {$\eta^{-1}(\ell)$};
\draw[orange!80!black,thick] (4.5,2.2) ellipse (0.36 and 0.12);
\draw[orange!80!black,thick,dashed] (4.5,1.35) ellipse (0.2 and 0.07);
\draw[orange!80!black,thick,dashed] (4.5,2.95) ellipse (0.52 and 0.17);
\node[font=\footnotesize,orange!80!black,align=left,anchor=west] at (5.15,2.25) {gauge orbit\\[-1pt]\& complexified};
\begin{scope}[xshift=8.3cm]
\node[font=\small] at (2.6,4.35) {\textbf{(b)} distributional averaging};
\foreach \x in {0.5,1.5,2.5,3.5,4.5}{\draw[gray!60] (\x,0.55) -- (\x,3.3); \draw[gray!60,dotted] (\x,0.05) -- (\x,0.55);}
\draw[thick,dashed,red!70!black] (0,-0.2) -- (5.0,-0.2) node[right,font=\footnotesize,align=left]{$\Pphy$\\[-1pt]base only};
\draw[thick,green!40!black] plot[smooth] coordinates {(0,1.25) (1.5,1.35) (2.5,1.2) (3.5,1.4) (4.9,1.3)};
\node[font=\footnotesize,green!40!black,anchor=west] at (5.0,1.3) {$\sigma_\chi$};
\draw[thick,green!40!black,dashed] plot[smooth] coordinates {(0,1.95) (1.5,2.05) (2.5,2.15) (3.5,1.95) (4.9,2.05)};
\node[font=\footnotesize,green!40!black,anchor=west] at (5.0,2.05) {$\sigma_*$};
\foreach \x in {0.5,1.5,2.5}{\draw[->,thick,blue!60!black] (\x+0.12,1.45) -- (\x+0.12,3.45);}
\node[font=\footnotesize,blue!60!black] at (1.5,3.7) {flow of $\Cq$};
\node[font=\footnotesize,gray!30!black,anchor=south] at (3.5,3.3) {$\eta_\Phi^{-1}(\ell)$};
\end{scope}
\end{tikzpicture}
\caption{The averaging map as a bundle. (a) When averaging is an orthogonal projection, the admissible rays form a vector bundle over $\Pphy$ whose zero section is the set of embedded exact solutions. Gauge orbits, and their complexifications, lie inside single fibres, which are in general much larger. The gradient flow of $\Cq=\ev{\Chat^2}$ runs down the fibres to the zero section. (b) When physical states are distributional, the representatives form an affine bundle with no zero section. A fixed seed profile $\chi$ gives a section $\sigma_\chi$, and the test-space inner product selects a canonical one, $\sigma_*$. The gradient flow still runs along the fibres, but there is nothing to reach in $\Ps$: it produces Weyl sequences of approximate solutions whose vector states accumulate on Dirac states.}
\label{fig:bundles}
\end{figure}
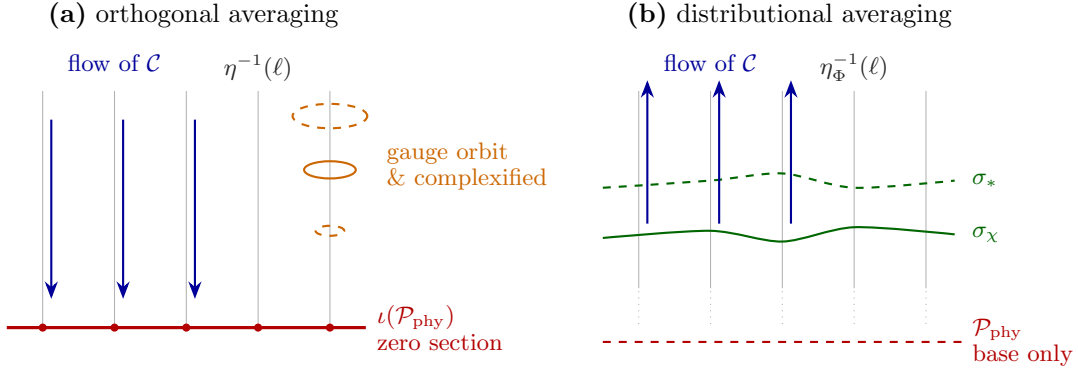

The physical state space has thus two simultaneous descriptions: the submanifold $\iota(\Pphy)\subset\Ps$, and the base of the averaging bundle. Every exact solution lies at the zero of its fibre. The homotopy $[u+k]\mapsto[u+sk]$, $0\le s\le1$, is a deformation retraction of $\Padm$ onto the zero section, so $\Padm$ is homotopy equivalent to $\Pphy$. This retraction describes the removal of the nonphysical components; it is neither a gauge transformation nor a unitary constraint evolution. Figure~\ref{fig:bundles} summarizes the picture, and contrasts it with the distributional case of Section~\ref{sec:clock}.

\subsection{Fibres, gauge orbits and complexified orbits}\label{sec:orbits}

Let us now compare the fibres of $\eta$ with the gauge orbits. Haar invariance gives $\Phat\,U(s)=\Phat$, so every gauge orbit in $\Padm$ lies in an averaging fibre. The converse fails: the fibre allows every change of the component in $\K$, whereas the group supplies only particular motions of it. Equation (\ref{eq:bundle}) is therefore a vector bundle of physical representatives, and it need not be a principal $G$-bundle; in general $\Padm/G\neq\Pphy$.

A three-dimensional example displays the geometry. Take $\Hil=\C^3$ and the dimensionless constraint $\Chat={\rm diag}(0,0,1)$. Averaging $U(\theta)=e^{-i\theta\Chat}$ over $U(1)$ gives $\Phat={\rm diag}(1,1,0)$ and
\begin{equation}\label{eq:exampleC3}
\eta:\CP^2\setminus\{[0:0:1]\}\longrightarrow\CP^1,\qquad [z_1:z_2:z_3]\longmapsto[z_1:z_2],
\end{equation}
the linear projection from the point $[0:0:1]$ \cite{GriffithsHarris}. The fibre above $[1:0]$ consists of the rays $[1:0:w]$, $w\in\C$, with normalized representatives
\begin{equation}\label{eq:exampleC3b}
|\Psi_w\rangle=\frac{|e_1\rangle+w\,|e_3\rangle}{\sqrt{1+|w|^2}},\qquad \Phat|\Psi_w\rangle=\frac{|e_1\rangle}{\sqrt{1+|w|^2}}.
\end{equation}
Every nonzero average normalizes to $|e_1\rangle$; the fibre records the amount and phase of the discarded component. The constraint flow sends $w\mapsto e^{-i\theta}w$, so for $w\neq0$ an orbit is a circle at fixed $|w|$, while at $w=0$ it is a point and the fibre is still a complex line. A radial parameter survives the quotient by these circles, which is why the group quotient is not the physical projective line. The omitted ray $[0:0:1]$, the limit $|w|\to\infty$, is the direction whose average vanishes. On the base charts $[1:v]$ and $[v^{-1}:1]$ the fibre coordinate transforms as $w'=w/v$. This is $L^*=\mathcal O(1)$ over $\CP^1$, which is nontrivial: no nonzero fibre coordinate can be chosen globally, although the prescription $w=0$, the zero section, is global.

What is the radial parameter? When the representation extends to the complexified group, as it does for compact $G$, every fibre of $\eta$ is $G_\C$-invariant \cite[Thm.~4.7]{Corichi2008}, so that
\begin{equation}\label{eq:chain}
\text{gauge orbits}\;\subset\;G_\C\text{-orbits}\;\subset\;\text{fibres of }\eta .
\end{equation}
In the example, $G_\C=\C^*$ acts on the fibre by $w\mapsto\lambda w$, so the punctured fibre is a single $G_\C$-orbit. The radial parameter that survives the quotient by circles is the imaginary part of the complexified gauge parameter. In general, however, the fibre is larger than the complexified orbits. In the model of Section~\ref{sec:oscillators}, for instance, $G_\C$ acts on $\K=\bigoplus_{n\neq N}\Hil_n$ by $\lambda^{n-N}$; its orbits have complex dimension one, while the fibre is infinite-dimensional. Both inclusions in (\ref{eq:chain}) are then strict. We shall see in Section~\ref{sec:clock} that the same chain persists in the distributional case, and in Section~\ref{sec:complexified} that the complexified directions have a precise classical meaning.

\subsection{The symplectic forms live on different spaces}

Define the retained norm fraction
\begin{equation}\label{eq:p}
p([\Psi])=\frac{\|\Phat\Psi\|^2}{\|\Psi\|^2}=\frac{1}{1+\|T\|^2},\qquad 0<p\le1,
\end{equation}
where $\|T\|=\|T(u)\|/\|u\|$ is the norm of $T$ on the line $\ell$.

\begin{proposition}\label{prop:forms}
On $\Padm$, $\eta^*\Omega_{\rm phy}=\Omega+i\hbar\,\partial\bar\partial\log p$. The degenerate directions of $\eta^*g_{\rm phy}$ and of $\eta^*\Omega_{\rm phy}$ are exactly $\ker{\rm d}\eta$. If $\K\neq0$, the ambient form $\Omega$ is not the pullback of any form on $\Pphy$.
\end{proposition}

\begin{proof}
If $\Psi(z)$ is a holomorphic local representative, so is $\Phat\Psi(z)$, and subtracting the potentials (\ref{eq:kahlerpot}) gives the formula. Since $\eta$ is a surjective submersion and $g_{\rm phy}$, $\Omega_{\rm phy}$ are nondegenerate, the pullbacks vanish on $X$ exactly when ${\rm d}\eta X=0$. For a nonzero vertical $V$, $JV$ is vertical because the fibres are complex submanifolds, and $\Omega(V,JV)=g(V,V)>0$, whereas a pulled-back form vanishes on this pair.
\end{proof}

Averaging is therefore not classical symplectic reduction of $\Padm$ with its ambient form. Reduction restricts to a constraint surface and removes its null directions \cite{MarsdenWeinstein}; averaging assigns a physical ray to seeds that generally are not solutions. In the example (\ref{eq:exampleC3}), $\Chat=1-\Phat$ gives $\Cq=1-p$, and on the fibre $[1:0:w]$ one has $h_{\rm FS}=|{\rm d}w|^2/(1+|w|^2)^2$, $\log p=-\log(1+|w|^2)$ and $c=\Cq=|w|^2/(1+|w|^2)$. The physical pullback vanishes on the fibre, consistently with Proposition~\ref{prop:forms}. Moving along the fibre is invisible after averaging, but it changes the quantum distance and the constraint expectations of the seed.

\subsection{The gradient flow of the squared constraint is vertical}\label{sec:vertical}

The flow that appeared in the heat-kernel realization of $\Phat$ has a simple description in the language of the bundle.

\begin{proposition}\label{prop:vertical-orth}
For $t\ge0$ the operator $e^{-t\Chat^2}$ maps $\Padm$ into itself and satisfies $\eta([e^{-t\Chat^2}\Psi])=\eta([\Psi])$. In the coordinates of Proposition~\ref{prop:bundle} it acts as $T\mapsto e^{-t\Chat^2}|_\K\circ T$, so it moves each admissible ray along its fibre and converges to the zero section as $t\to\infty$, exponentially fast when $0$ is isolated in $\spec\Chat^2$.
\end{proposition}

\begin{proof}
$e^{-t\Chat^2}$ commutes with $\Phat$ and is the identity on $\Hphy$, so $\Phat e^{-t\Chat^2}\Psi=\Phat\Psi\neq0$ and $e^{-t\Chat^2}(u+k)=u+e^{-t\Chat^2}k$, with $e^{-t\Chat^2}k\in\K$. By the spectral theorem $e^{-t\Chat^2}k\to0$ for $k\in\K$, with rate $e^{-t\Delta}$ when the spectrum of $\Chat^2$ on $\K$ is bounded below by $\Delta>0$.
\end{proof}

By \vthree, the rays $[e^{-t\Chat^2}\Psi]$ are, up to a rescaling of the parameter, the negative gradient flow of $\Cq$ for the metric $g$. The flow is thus \emph{vertical} for the averaging bundle: it changes the seed but never the physical state. In the example (\ref{eq:exampleC3}) it acts on the fibre as $w\mapsto e^{-t}w$, the imaginary part of the complexified gauge flow. This is the precise sense in which heat-kernel averaging and group averaging produce the same physical state: the former moves along the fibre, the latter jumps to its origin.

\subsection{Observables}

A self-adjoint physical operator $\Ohat_{\rm phy}$ defines on $\Padm$ the function
\begin{equation}\label{eq:otilde}
\tilde o([\Psi])=\frac{\ip{\Phat\Psi}{\Ohat_{\rm phy}\Phat\Psi}}{\|\Phat\Psi\|^2},
\end{equation}
wherever $\Phat\Psi$ lies in its domain. It is constant on every fibre, so physical observables are functions on the base. Which kinematical flows descend to the base? The answer is the following.

\begin{proposition}\label{prop:descent}
Let $\Ohat$ be self-adjoint. The following are equivalent:
\begin{enumerate}[(a)]
\item $e^{-it\Ohat/\hbar}$ maps $\Padm$ into itself for every $t\in\R$;
\item $\Ohat$ commutes strongly with $\Phat$, i.e.\ $\Phat\Dom\Ohat\subset\Dom\Ohat$ and $\Phat\Ohat\subset\Ohat\Phat$.
\end{enumerate}
When they hold, $\eta\circ e^{-it\Ohat/\hbar}=e^{-it\Ohat_{\rm phy}/\hbar}\circ\eta$, where $\Ohat_{\rm phy}$ is the restriction of $\Ohat$ to $\Hphy$.
\end{proposition}

\begin{proof}
If (b) holds, $[\Phat,e^{-it\Ohat/\hbar}]=0$, so $\Phat e^{-it\Ohat/\hbar}\Psi=e^{-it\Ohat_{\rm phy}/\hbar}\Phat\Psi\neq0$ whenever $\Phat\Psi\neq0$; this gives (a) and the intertwining relation. Conversely, write $U_t=e^{-it\Ohat/\hbar}$. Applying (a) to $t$ and to $-t$ shows that the ray map of $U_t$ restricts to a bijection of $\Padm$, hence of its complement $\PP(\K)$. A unitary operator that maps every ray of $\K$ into $\PP(\K)$ maps $\K$ into $\K$; applied to $U_{\pm t}$ this gives $U_t\K=\K$, so $U_t$ also preserves $\Hphy=\K^\perp$. Hence $[U_t,\Phat]=0$ for all $t$, which is (b).
\end{proof}

For bounded $\Ohat$, commuting with $\Phat$ is equivalent to $\Ohat\Hphy\subset\Hphy$, which by Proposition~\ref{prop:hessian} is the Hessian criterion of the embedded description. The two descriptions therefore select the same observable motions, and the chain
\[
\text{the flow descends}\iff[\Ohat,\Phat]=0\iff\text{the Hessian of }\Cq\text{ vanishes along the flow on }\Pphy
\]
holds, the last equivalence for bounded $\Ohat$. Proposition~\ref{prop:descent} also covers unbounded observables, such as the Schwinger operators of Section~\ref{sec:oscillators}. Let us note, however, that commuting with the constraint does not make kinematical expectations constant along fibres. In the example (\ref{eq:exampleC3}), $\Ohat={\rm diag}(1,-1,7)$ commutes with $\Chat$, yet on $[1:0:w]$ its kinematical expectation is $(1+7|w|^2)/(1+|w|^2)$, while $\tilde o=1$. A fibre records the equality of the resulting physical state, not of all kinematical predictions.

\section{Two oscillators with fixed total energy}\label{sec:oscillators}

The fixed-total-energy model of \cite[Sec.~V.C]{ABC} is the simplest system in which the full bundle of seeds and the bundle of a preferred family of seeds can be compared. It also shows how classical gauge orbits sit inside the averaging fibres.

\paragraph{Exact solutions.} For two oscillators of frequency $\nu$, take
\begin{equation}\label{eq:osc-constraint}
\Chat=\widehat H_1+\widehat H_2-E=\hbar\nu\,(\widehat N_1+\widehat N_2-N),\qquad E=\hbar\nu(N+1),
\end{equation}
with $\widehat H_i=\hbar\nu(\widehat N_i+\frac12)$ and an integer $N\ge0$.\footnote{This is the operator $\hbar\nu(\hat n+1)-E$ used in \cite{CorichiGauge}, with $\hat n=\widehat N_1+\widehat N_2$ and $E=\hbar\nu(N+1)$.} The dimensionless constraint $\widehat N_1+\widehat N_2-N$ is the normal-ordered form of \cite{ABC}, with their integer parameter identified with $N$. Averaging over $U(1)$,
\begin{equation}\label{eq:osc-P}
\Phat=\frac1{2\pi}\int_0^{2\pi}e^{-i\theta(\widehat N_1+\widehat N_2-N)}\,{\rm d}\theta,\qquad \Hphy={\rm span}\{|n,N-n\rangle:0\le n\le N\},
\end{equation}
so $\Pphy=\CP^N$ is the space of all exact pure physical states, embedded in the Fock ray space. The full averaging bundle is (\ref{eq:bundle}) with infinite-dimensional $\K=\Hphy^\perp$, hence topologically trivial. Its fibres allow arbitrary components at all other occupation levels, whose phases the constraint group rotates.

\paragraph{Coherent seeds.} Let $|\alpha))=e^{\alpha_1a_1^\dagger+\alpha_2a_2^\dagger}|0,0\rangle$, $R=|\alpha_1|^2+|\alpha_2|^2$, and $|\Psi_\alpha\rangle=e^{-R/2}|\alpha))$. Averaging extracts the homogeneous part of degree $N$,
\begin{equation}\label{eq:vN}
v_N(\alpha)=\Phat|\alpha))=\sum_{n=0}^N\frac{\alpha_1^n\alpha_2^{N-n}}{\sqrt{n!\,(N-n)!}}\,|n,N-n\rangle,\qquad \|v_N(\alpha)\|^2=\frac{R^N}{N!},
\end{equation}
and, for $\beta=\alpha/\sqrt R$, the normalized physical state is
\begin{equation}\label{eq:spincoh}
|\psi_N(\beta)\rangle=\frac{(\beta_1a_1^\dagger+\beta_2a_2^\dagger)^N}{\sqrt{N!}}\,|0,0\rangle,\qquad |\beta_1|^2+|\beta_2|^2=1,
\end{equation}
an SU(2) spin coherent state of spin $N/2$ \cite{Radcliffe,ACGT,Perelomov}. The norm and common phase of $\alpha$ do not affect the physical ray, although the phase of the label changes the kinematical ray. Since $v_N(\lambda\alpha)=\lambda^Nv_N(\alpha)$, and the coefficient ratio of $|N-1,1\rangle$ to $|N,0\rangle$ is $\sqrt N\alpha_2/\alpha_1$, two coherent seeds give the same physical ray exactly when their labels are proportional. For $N\ge1$ the restricted map factors as
\begin{equation}\label{eq:veronese}
\C^2\setminus\{0\}\longrightarrow\CP^1\xrightarrow{\;\nu_N\;}\CP^N ,
\end{equation}
where the first map is $\mathcal O(-1)$ with its zero section removed, a principal $\C^*$ bundle. The map $\nu_N$ is the degree-$N$ Veronese embedding \cite{GriffithsHarris} composed with the diagonal rescaling by $(n!\,(N-n)!)^{-1/2}$, and its image is the orbit of spin coherent states. For $N=0$ every seed averages to the vacuum, and for $N=1$ the coherent image is all of $\Pphy$. For $N=2$, the coordinates $[w_0:w_1:w_2]=[\alpha_1^2/\sqrt2:\alpha_1\alpha_2:\alpha_2^2/\sqrt2]$ satisfy $w_1^2=2w_0w_2$, a conic in $\CP^2$ that misses the exact physical ray $|1,1\rangle$. For $N>1$ the coherent image has complex dimension one in the $N$-dimensional $\Pphy$: not every physical state has a representative in a prescribed seed family, although every physical state is embedded in $\Hil$.

There is a simple way to understand the fibres of (\ref{eq:veronese}). The complexified constraint group acts on coherent seeds by $e^{-i\theta(\widehat N_1+\widehat N_2-N)}|\alpha))=e^{iN\theta}|e^{-i\theta}\alpha))$, so as $\theta$ ranges over $\C$ the label $\alpha$ ranges over $\C^*\alpha$. Within the coherent family, therefore, \emph{each averaging fibre is exactly one complexified gauge orbit}. Its real part, $|e^{-i\theta}|=1$, is the common-phase circle of $\alpha$; its imaginary part is the radial direction $R\mapsto e^{2\,{\rm Im}\,\theta}R$. In the full bundle, by contrast, the fibres are infinite-dimensional and both inclusions in (\ref{eq:chain}) are strict.

\paragraph{Geometry and lifting.} On the chart $\alpha=(1,z)$,
\begin{equation}\label{eq:veronese-form}
\nu_N^*\Omega_{\rm phy}=N\,i\hbar\,\frac{{\rm d}z\wedge{\rm d}\bar z}{(1+|z|^2)^2},
\end{equation}
which is $N$ times the quantum symplectic form of a projective line, while on the seed manifold the ambient form is $i\hbar\sum_i{\rm d}\alpha_i\wedge{\rm d}\bar\alpha_i$, from $\log\||\alpha))\|^2=R$. The seed's constraint data are
\begin{equation}\label{eq:osc-data}
p=e^{-R}\frac{R^N}{N!},\qquad c=\hbar\nu(R-N),\qquad \Cq=(\hbar\nu)^2\bigl(R+(R-N)^2\bigr),
\end{equation}
since the total occupation is Poisson distributed with mean and variance $R$. At $R=N$, $p$ is maximal and $c=0$, but $\Cq=(\hbar\nu)^2N>0$: the seed is centred on the constraint in expectation and is still not a solution. The excess is precisely the orbit-speed term of (\ref{eq:decomp}).

Restricting to $R=N$ gives a three-sphere, whose quotient by the common phase is the Hopf fibration $S^3\to\CP^1$, as noted in \cite{ABC}. This is the classical reduction of the excitation constraint, and its reduced form agrees with (\ref{eq:veronese-form}). The reduced sphere with prequantum bundle $\mathcal O(N)$ quantizes to a space of dimension $N+1=\dim\Hphy$, an instance of the principle that quantization commutes with reduction \cite{GuilleminSternberg}. A radius can be fixed globally, but a common phase cannot: the local choices $\alpha_0=\sqrt N(1,z)/\sqrt{1+|z|^2}$ and $\alpha_1=\sqrt N(\zeta,1)/\sqrt{1+|\zeta|^2}$, $\zeta=1/z$, satisfy $\alpha_1=(|z|/z)\,\alpha_0$ on the overlap. This transition function has nonzero winding, which is the obstruction to a global section of the Hopf bundle. Such local coherent lifts select kinematical seeds, not exact solutions, so they do not conflict with the zero section $\iota$.

\paragraph{Dirac observables and fluctuations.} The Schwinger operators
\begin{equation}\label{eq:schwinger}
\widehat J_1=\tfrac12\bigl(a_1^\dagger a_2+a_2^\dagger a_1\bigr),\qquad \widehat J_2=\tfrac1{2i}\bigl(a_1^\dagger a_2-a_2^\dagger a_1\bigr),\qquad \widehat J_3=\tfrac12\bigl(\widehat N_1-\widehat N_2\bigr)
\end{equation}
commute strongly with $\Chat$, satisfy $[\widehat J_i,\widehat J_j]=i\epsilon_{ijk}\widehat J_k$, and act on $\Hphy$ as the spin-$N/2$ representation. For $\mathbf n=(2\,{\rm Re}\,\bar\beta_1\beta_2,\,2\,{\rm Im}\,\bar\beta_1\beta_2,\,|\beta_1|^2-|\beta_2|^2)$,
\begin{equation}\label{eq:moments}
\ev{\widehat J_i}_{\rm phy}=\frac N2\,n_i,\qquad (\Delta J_i)^2_{\rm phy}=\frac N4\bigl(1-n_i^2\bigr),\qquad \ev{\widehat J_i}_{\rm kin}=\frac R2\,n_i,\qquad (\Delta J_i)^2_{\rm kin}=\frac R4 .
\end{equation}
The physical values are the standard spin-coherent moments \cite{ACGT}, and the kinematical ones follow by normal ordering. At $R=N$ the means coincide, while averaging reduces each variance by $Nn_i^2/4=\ev{\widehat J_i}^2_{\rm phy}/N$. This reproduces, in spin language, the fluctuation comparison of \cite{ABC}, whose observables obey the same angular-momentum algebra. Along a coherent averaging fibre, changing $R$ alters the kinematical moments but none of the physical predictions.

\section{A clock constraint and distributional physical states}\label{sec:clock}

What survives when exact solutions are not normalizable? A parametrized oscillator answers this question without the complications of field theory, and it is the simplest model of a Hamiltonian constraint with an ideal internal clock.

\subsection{The model}

The classical constraint $p_t+H_{\rm osc}$ treats the clock reading $t$ as a canonical variable. Quantum mechanically,
\begin{equation}\label{eq:clock}
\Hil=L^2(\R,{\rm d}t)\otimes\Hil_{\rm osc},\qquad \Chat=\hat p_t\otimes1+1\otimes\widehat H_{\rm osc},\qquad \widehat H_{\rm osc}|n\rangle=E_n|n\rangle,
\end{equation}
with $\hat p_t=-i\hbar\partial_t$ and $\Chat$ the closure of the sum on the algebraic tensor product. The constraint equation $\Chat\Psi=0$ is $i\hbar\partial_t\Psi=\widehat H_{\rm osc}\Psi$, with generalized solutions
\begin{equation}\label{eq:histories}
\Psi_f(t)=\sum_{n\ge0}f_n\,e^{-iE_nt/\hbar}|n\rangle,\qquad \Hphy\simeq\ell^2,\qquad \ip{f}{f'}_{\rm phy}=\sum_n\bar f_nf'_n .
\end{equation}
Their oscillator norm is constant in $t$, so every nonzero history has infinite kinematical norm; in $p_t$-space it is supported on the measure-zero set $p_t=-E_n$. Hence $\ker\Chat=\{0\}$, and by the mean ergodic theorem the Ces\`aro averages of $e^{-it\Chat/\hbar}$ converge strongly to zero. In the Hilbert topology, averaging annihilates every kinematical vector. A nontrivial physical space nevertheless exists in a weaker topology. That the induced inner product reproduces $\ell^2$ in this model is standard \cite{HartleMarolf}; here we use it to analyse the seed-to-state map. We assume, as for the oscillator, that $\widehat H_{\rm osc}$ has discrete spectrum. Degeneracies are harmless, whereas continuous spectrum would require a direct integral.

\subsection{Averaging as evaluation on the constraint shell}

Fix an energy scale $\eps>0$, and let $\tilde\Psi(p)=(2\pi\hbar)^{-1/2}\int e^{-ipt/\hbar}\Psi(t)\,{\rm d}t$. The map $\Psi\mapsto\phi$, $\phi_n(x)=\sqrt\eps\,\langle n|\tilde\Psi(\eps x)\rangle$, is unitary from $\Hil$ onto $L^2(\R,{\rm d}x;\ell^2)$, and $\Chat$ acts on $\phi_n$ as multiplication by $\eps(x+e_n)$, with $e_n=E_n/\eps$. The unitary shift
\begin{equation}\label{eq:shift}
(W\phi)_n(y)=\phi_n(y-e_n),\qquad W\Chat W^{-1}=\eps y,
\end{equation}
moves every constraint shell to $y=0$; we write $\tilde\phi=W\phi$. Take the dense test space
\begin{equation}\label{eq:testspace}
\Phi=W^{-1}H^1(\R;\ell^2),\qquad \|\phi\|^2_\Phi=\int_\R\bigl(\|\tilde\phi(y)\|^2+\|\tilde\phi'(y)\|^2\bigr)\,{\rm d}y ,
\end{equation}
continuously embedded in $\Hil$. For $\tilde\phi\in H^1$, integrating $\frac{{\rm d}}{{\rm d}y}\|\tilde\phi\|^2$ over each half-line and using the Cauchy--Schwarz inequality gives
\begin{equation}\label{eq:trace}
\|\tilde\phi(0)\|^2\le\|\tilde\phi\|_{L^2}\,\|\tilde\phi'\|_{L^2}\le\tfrac12\|\phi\|^2_\Phi ,
\end{equation}
so evaluation at the shell is continuous on $\Phi$. Multiplication by $e^{-i\lambda y}$ maps $H^1$ continuously into itself, so the constraint group $\{e^{-i\lambda\Chat/\eps}\}_{\lambda\in\R}$ preserves $\Phi$.

\begin{proposition}\label{prop:clock-raq}
On the dense core $\D=W^{-1}(\mathcal S(\R)\otimes_{\rm alg}c_{00})$, with $c_{00}$ the finite sequences, the group-averaging integral converges absolutely and
\begin{equation}\label{eq:B}
B(\phi,\psi)=\frac1{2\pi}\int_\R\ip{\phi}{e^{-i\lambda\Chat/\eps}\psi}\,{\rm d}\lambda=\ip{Q\phi}{Q\psi}_{\rm phy},\qquad (Q\phi)_n=\tilde\phi_n(0)=\phi_n(-e_n).
\end{equation}
$B$ extends uniquely to a continuous positive form on $\Phi$, given by the same expression, with null space $\N=\ker Q$. Any approximate-identity regularization of the $\lambda$-integral, for instance the factor $e^{-\tau\lambda^2}$ with $\tau\downarrow0$, gives the same form on all of $\Phi$. Moreover, $Q\Chat=0$ on $\D$, so $f\mapsto\ip{f}{Q(\cdot)}_{\rm phy}$ are distributional solutions of the constraint.
\end{proposition}

\begin{proof}
On $\D$, $G(y)=\langle\tilde\phi(y),\tilde\psi(y)\rangle$ is a Schwartz function, and $\ip{\phi}{e^{-i\lambda\Chat/\eps}\psi}=\int e^{-i\lambda y}G(y)\,{\rm d}y$. Fourier inversion gives $G(0)$. Continuity of $\phi\mapsto Q\phi$ in the $\Phi$ topology, (\ref{eq:trace}), gives the extension. With a regulator, the $\lambda$-integral produces the kernel $G_\tau(y)=e^{-y^2/4\tau}/\sqrt{4\pi\tau}$, an approximate identity acting on the bounded continuous integrable function $G$, whose limit is again $G(0)$. Finally, $(W\Chat\phi)(0)=\eps\,y\,\tilde\phi(y)|_{y=0}=0$.
\end{proof}

Without a regulator the integral need not converge absolutely on all of $\Phi$, since $G\in W^{1,1}$ only. Defining $B$ on $\D$ and extending it by continuity meets the requirements of refined algebraic quantization \cite{GiuliniMarolf1,GiuliniMarolf2}. The test space is a Hilbert space and need not be nuclear. The Gaussian regulator has a geometric meaning. Since the $\lambda$-integral of $e^{-\tau\lambda^2}e^{-i\lambda y}$ is the heat kernel,
\begin{equation}\label{eq:regulated}
\frac1{2\pi}\int_\R e^{-\tau\lambda^2}\ip{\phi}{e^{-i\lambda\Chat/\eps}\psi}\,{\rm d}\lambda=\frac{1}{\sqrt{4\pi\tau}}\,\ip{\phi}{e^{-\Chat^2/(4\tau\eps^2)}\psi},
\end{equation}
the regulated average is the kinematical overlap along the gradient flow of $\Cq$ at time $1/(4\tau\eps^2)$, suitably rescaled, and Proposition~\ref{prop:clock-raq} is the clock-model instance of the rigging-map formula of \cite[Thm.~7.3]{Corichi2008}.

The stronger topology is essential. For $\chi\in\mathcal S(\R)$ with $\chi(0)=1$, the vectors $\tilde\phi^\delta(y)=\chi(y/\delta)f$ have $Q\phi^\delta=f$, while $\|\phi^\delta\|^2_\Hil=\delta\,\|\chi\|^2_{L^2}\|f\|^2\to0$. Hence $Q$ has no continuous extension to $\Hil$.

\subsection{The affine bundle and its sections}

A profile $\chi\in H^1(\R)$ with $\chi(0)=1$ defines a continuous right inverse of $Q$,
\begin{equation}\label{eq:sigma}
(\sigma_\chi f)_n(x)=\chi(x+e_n)\,f_n,\qquad W\sigma_\chi f=\chi\otimes f,\qquad Q\sigma_\chi={\rm id},\qquad \Phi=\sigma_\chi\Hphy\oplus\N ,
\end{equation}
with $\|\sigma_\chi f\|_\Phi=\|\chi\|_{H^1}\|f\|$. Every $f\in\ell^2$ has a lift, so the quotient $\Phi/\N$ already exhausts the completed physical Hilbert space.

\begin{proposition}\label{prop:affine}
The map $\eta_\Phi:\PP(\Phi)\setminus\PP(\N)\to\Pphy$, $[\phi]\mapsto[Q\phi]$, is a holomorphic affine bundle modelled on $L^*\otimes\underline\N$. Above a physical line $\ell$, the fibre is the set of linear right inverses $r:\ell\to\Phi$ of $Q$, an affine space over $\Hom(\ell,\N)$. Each profile $\chi$ gives a global section $[f]\mapsto[\sigma_\chi f]$, and hence an identification with the vector bundle $\Tot(L^*\otimes\underline\N)$. Changing the profile translates the origin of each fibre by $(\sigma_{\chi'}-\sigma_\chi)|_\ell$.
\end{proposition}

\begin{proof}
A seed $\phi$ with $Q\phi\neq0$ defines $r:\ell=\C Q\phi\to\Phi$ by $r(Q\phi)=\phi$, and two right inverses differ by an element of $\Hom(\ell,\N)$. For charts, choose a bounded functional $\varpi$ on $\Hphy$, normalize $\varpi(f)=1$, and write seeds as $\sigma_\chi f+k$, $k\in\N$. These are product charts in the $\Phi$ topology, with the transition functions of Proposition~\ref{prop:bundle}.
\end{proof}

The contrast with Section~\ref{sec:bundle} is the absence of a distinguished zero section: the would-be origin of each fibre is an exact solution, which is not in $\Phi$. Since $\N$ is infinite-dimensional, the associated vector bundle is topologically trivial. Constraint evolution multiplies $\tilde\phi(y)$ by the $y$-dependent phase $e^{-i\lambda y}$. It preserves $|\tilde\phi(y)|$ and the shell value, so two profiles with different moduli lie in the same fibre but are not related by any constraint motion. The complexified flow, where defined, multiplies $\tilde\phi(y)$ by $e^{\tau y}$. It fixes the shell value and changes the modulus of the profile, mapping a Gaussian to a shifted Gaussian, but it cannot relate, for example, $e^{-y^2/2}$ and $e^{-|y|}$. The chain (\ref{eq:chain}) thus persists, with both inclusions strict.

\paragraph{A canonical section.} Because $\Phi$ is a Hilbert space and $\N$ is closed, the $\Phi$-orthogonal complement of $\N$ gives a splitting fixed by the test-space inner product. The minimal-norm lift with $\tilde\phi(0)=f$ solves $(1-\partial_y^2)\tilde\phi\propto\delta(y)f$, so
\begin{equation}\label{eq:canonical-clock}
\chi_*(y)=e^{-|y|},\qquad \|\sigma_{\chi_*}f\|^2_\Phi=2\|f\|^2 ,
\end{equation}
which saturates the trace inequality (\ref{eq:trace}). This section is canonical given $(\Phi,\|\cdot\|_\Phi)$, but it depends on the scale $\eps$ through the variable $y=(p_t+E_n)/\eps$. Different $\eps$ give equivalent norms, so the space $\Phi$ and its topology are $\eps$-independent, while the minimizer is not. Nothing in $\Chat$ and the kinematical inner product alone selects a section.

\paragraph{Profile lifts are isometric embeddings of non-solutions.} Since $\|\sigma_\chi f\|_\Hil=\|\chi\|_{L^2}\|f\|$ and $\ip{\sigma_\chi f}{\sigma_\chi f'}_\Hil=\|\chi\|^2_{L^2}\ip{f}{f'}$, the induced map $j_\chi:\Pphy\to\Ps$, $[f]\mapsto[\sigma_\chi f]$, is a holomorphic K\"ahler-isometric embedding: the constant factor drops out of (\ref{eq:kahlerpot}). Its image contains no exact solution, since $\ker\Chat=\{0\}$. On the image,
\begin{equation}\label{eq:lift-moments}
\ev{\Chat}=\eps\,\frac{\int y|\chi|^2{\rm d}y}{\int|\chi|^2{\rm d}y},\qquad \Cq=\ev{\Chat^2}=\eps^2\,\frac{\int y^2|\chi|^2{\rm d}y}{\int|\chi|^2{\rm d}y},
\end{equation}
independently of $f$. For the Gaussian $e^{-y^2/2}$ and for the canonical profile (\ref{eq:canonical-clock}) alike, $\ev{\Chat}=0$ and $\Cq=\eps^2/2$. Let us stress that the isometry depends on the profile being the same for every oscillator level. A level-dependent profile $\chi_n$ gives instead the Hermitian form $\langle f,{\rm diag}(\|\chi_n\|^2_{L^2})f\rangle$, and the induced embedding is K\"ahler isometric only if $\|\chi_n\|_{L^2}$ is independent of $n$. This is precisely the mechanism that will operate in Section~\ref{sec:slqc}.

\paragraph{Weyl sequences and the gradient flow.} Narrowing the profile, $\chi(y)\mapsto\chi(y/\delta)$, keeps $Q\sigma_\chi f=f$ and gives $\Cq\propto\eps^2\delta^2\to0$, while the normalized lifts converge weakly to zero. The lifts of a fixed physical state thus form Weyl sequences: approximate solutions for the value $0$ in the continuous spectrum of $\Chat$, without a limiting ray. This connects the distributional construction with the function $\Cq$ of Section~\ref{sec:embedded}, and with Klauder's projections onto $|\Chat|\le\delta$ \cite{Klauder1,Klauder2}. The narrowing is not an ad hoc device: it is generated by the same flow that, in the orthogonal case, ran down the fibres.

\begin{proposition}\label{prop:vertical}
The operators $e^{-t\Chat^2}$, $t\ge0$, preserve $\Phi$ and satisfy $Q\,e^{-t\Chat^2}=Q$; they therefore act on the fibres of $\eta_\Phi$. Starting from $\sigma_\chi f$, they produce the profiles $\chi(y)\,e^{-t\eps^2y^2}$, whose normalized lifts form a Weyl sequence as $t\to\infty$. Every weak-$*$ cluster point of the corresponding vector states is a Dirac state.
\end{proposition}

\begin{proof}
By (\ref{eq:shift}), $We^{-t\Chat^2}W^{-1}$ is multiplication by $e^{-t\eps^2y^2}$, which maps $H^1$ to itself and equals $1$ at $y=0$. The spectral measure of $\Chat$ in the state $\sigma_\chi f$ has density proportional to $|\chi(y)|^2\|f\|^2$ in the variable $y$, which is positive near $y=0$. Along the normalized flow this weight concentrates at $0$, which is not an eigenvalue, so $\Cq\to0$ and the vectors converge weakly to zero. The last statement is \cite[Thm.~7.2]{Corichi2008}.
\end{proof}

In both regimes, therefore, the gradient flow of $\Cq$ is vertical for the averaging bundle. In the orthogonal case it reaches the zero section; in the distributional case there is no zero section in $\Ps$, and the flow escapes to the Dirac states `at infinity' of \cite{GrundlingHurst}, while the physical state it represents never changes (Figure~\ref{fig:bundles}).

\subsection{Observables and clock-conditioned states}

For a bounded self-adjoint $\widehat A$ on $\Hphy$, define
\begin{equation}\label{eq:OA}
\Ohat_A=W^{-1}(1\otimes\widehat A)W,\qquad (\Ohat_A\phi)_n(x)=\sum_mA_{nm}\,\phi_m(x+e_n-e_m),\qquad Q\,\Ohat_A=\widehat A\,Q .
\end{equation}
This kinematical operator mixes constraint shells and is nonlocal in $p_t$. It preserves $\Phi$ and $\N$, commutes strongly with $\Chat=W^{-1}\eps yW$, maps fibres to fibres, and satisfies $\Ohat_A\sigma_\chi=\sigma_\chi\widehat A$, so every profile section carries the physical motion. Its physical expectation $\ip{Q\phi}{\widehat AQ\phi}/\|Q\phi\|^2$ does not depend on the seed or the profile. Intertwining with $Q$ is the distributional replacement for the Hessian criterion of Proposition~\ref{prop:hessian}.

Evaluating the history (\ref{eq:histories}) at the clock reading $\tau$ gives the oscillator state
\begin{equation}\label{eq:PW}
{\rm ev}_\tau f=e^{-i\widehat H_{\rm osc}\tau/\hbar}f,\qquad \ip{{\rm ev}_\tau f}{\widehat A\,{\rm ev}_\tau f}=\ip{f}{\widehat A(\tau)f},\qquad \widehat A(\tau)=e^{i\widehat H_{\rm osc}\tau/\hbar}\widehat Ae^{-i\widehat H_{\rm osc}\tau/\hbar}.
\end{equation}
This is the Page--Wootters conditional state \cite{PageWootters}, and $\Ohat_{A(\tau)}$ is the corresponding Dirac observable: Rovelli's evolving constant of motion, or complete observable \cite{Rovelli,Dittrich}. Their agreement is an instance of the equivalence established in \cite{HSL1}. For example, if $\widehat A$ projects onto $(|0\rangle+|1\rangle)/\sqrt2$, the probability $|f_0e^{-iE_0\tau/\hbar}+f_1e^{-iE_1\tau/\hbar}|^2/(2\|f\|^2)$ varies with $\tau$ and does not depend on which seed has $Q\phi=f$. Three objects must be kept apart:
\begin{itemize}[itemsep=0pt]
\item ${\rm ev}_\tau f\in\Hil_{\rm osc}$ is a clock-conditioned state;
\item $\sigma_\chi f\in\Hil$ is a normalizable seed;
\item the history $\Psi_f$ is a generalized solution.
\end{itemize}
A clock slice is not a section of the averaging bundle. Changing $\tau$ changes the relational question asked of a physical state, whereas changing $\chi$ changes a section and leaves every Dirac prediction unchanged.

\section{Solvable loop quantum cosmology}\label{sec:slqc}

The clock model has an ideal clock: $\hat p_t$ has spectrum $\R$, so at fixed oscillator label the constraint shell lies in the interior of a spectrum unbounded on both sides. A Hamiltonian constraint in gravity is typically quadratic in the clock momentum. Solvable loop quantum cosmology (sLQC) \cite{ACS,APS,AshtekarSingh} is the simplest model in which this matters, and in which group averaging is fully explicit \cite{KLP}. Let us now see which of the structures of Section~\ref{sec:clock} survive, and what changes.

\subsection{The model}

For a spatially flat isotropic universe with a massless scalar field $\phi$ as internal clock, the kinematical Hilbert space is $\Hil=L^2(\R,{\rm d}\phi)\otimes\Hil_{\rm grav}$, and the quantum constraint reads
\begin{equation}\label{eq:slqc}
\partial_\phi^2\chi=-\Theta\chi,\qquad \Chat=\hat p_\phi^2-\Theta,\qquad \hat p_\phi=-i\partial_\phi ,
\end{equation}
in the units of \cite{ACS}. In the variable $x=(12\pi G)^{-1/2}\ln\tan(\lambda b/2)$, the gravitational operator is $\Theta=-\partial_x^2$, and states are antisymmetric, $\chi(-x,\phi)=-\chi(x,\phi)$ \cite{ACS}. We use only the spectral structure: on the antisymmetric sector, $\Theta\ge0$ has simple absolutely continuous spectrum, with spectral representation $\Hil_{\rm grav}\simeq L^2(\R_+,{\rm d}k)$ in which $\Theta$ acts as $k^2$. The physical states of \cite{ACS} are the positive-frequency solutions, $-i\partial_\phi\chi=\sqrt\Theta\,\chi$, with physical inner product $2\int{\rm d}k\,|k|\,\overline{\tilde\chi_1(k)}\,\tilde\chi_2(k)$, a Klein--Gordon product. Since the joint spectral density of a solution is supported on the shells $p_\phi=\pm k$, a set of measure zero, $\ker\Chat=\{0\}$. By the mean ergodic theorem, Ces\`aro averages of $e^{-i\lambda\Chat}$ therefore converge strongly to zero, exactly as in the clock model.

\subsection{Shell coordinates and the spectral edge}

Let $\psi(p,k)$ denote a kinematical state in the joint spectral representation of $\hat p_\phi$ and $\Theta$. At fixed $k$, the constraint takes the value $c=p^2-k^2\in[-k^2,\infty)$ on each of the two sheets $p=\sigma\sqrt{c+k^2}$, $\sigma=\pm$. Fix a momentum scale $\eps>0$ and set
\begin{equation}\label{eq:shellcoords}
y=\frac{c}{\eps^2},\qquad \kappa_k=\frac{k^2}{\eps^2},\qquad \tilde\psi_\sigma(y,k)=\eps\,\frac{\psi\bigl(\sigma\sqrt{\eps^2y+k^2},\,k\bigr)}{\bigl(2\sqrt{\eps^2y+k^2}\bigr)^{1/2}} .
\end{equation}
The map $\psi\mapsto(\tilde\psi_+,\tilde\psi_-)$ is unitary from $\Hil$ onto $\bigoplus_\sigma\int^\oplus L^2((-\kappa_k,\infty),{\rm d}y)\,{\rm d}k$, and $\Chat$ acts as multiplication by $\eps^2y$. The shell is $y=0$ on both sheets, as in (\ref{eq:shift}), but now the range of $y$ at fixed $k$ is the half-line $(-\kappa_k,\infty)$: the clock Hamiltonian $\hat p_\phi^2$ is bounded below, so the constraint spectrum at fixed $k$ has an edge at distance $\kappa_k$ below the shell. The edge is the point $p_\phi=0$, where the two sheets meet. This spectral edge, absent for an ideal clock, controls everything that follows. Since the bounce volume grows linearly with $|p_\phi|$ \cite{ACS}, and $|p_\phi|=k$ on the shell, small $\kappa_k$ corresponds to universes that are small on the scale set by $\eps$. Let us emphasize from the outset that $\eps$ is a scale introduced by the test space, which nothing in sLQC fixes; we shall return to this point in Proposition~\ref{prop:weights}.

\subsection{Group averaging on an explicit test space}

Take
\begin{equation}\label{eq:slqc-test}
\Phi=\Bigl\{\psi:\ \|\psi\|^2_\Phi=\sum_\sigma\int_0^\infty{\rm d}k\int_{-\kappa_k}^\infty\bigl(|\tilde\psi_\sigma|^2+|\partial_y\tilde\psi_\sigma|^2\bigr){\rm d}y<\infty\Bigr\}.
\end{equation}
Using only the half-line $y\ge0$, which is available for every $k$, one obtains the uniform trace bound $|\tilde\psi_\sigma(0,k)|^2\le\|\tilde\psi_\sigma(\cdot,k)\|^2_{H^1(-\kappa_k,\infty)}$. Hence $Q\psi=(\tilde\psi_+(0,\cdot),\tilde\psi_-(0,\cdot))$ is continuous on $\Phi$. The constraint group acts by multiplication by $e^{-i\lambda y}$ and preserves $\Phi$.

\begin{proposition}\label{prop:slqc-raq}
Let $\D\subset\Phi$ consist of the $\psi$ whose $\tilde\psi_\sigma$ are smooth with compact support in $\{(y,k):k>0,\ y>-\kappa_k\}$. On $\D$ the averaging integral converges absolutely, and
\begin{equation}\label{eq:slqc-B}
B(\phi,\psi)=\frac1{2\pi}\int_\R\ip{\phi}{e^{-i\lambda\Chat/\eps^2}\psi}\,{\rm d}\lambda=\sum_\sigma\int_0^\infty\overline{\tilde\phi_\sigma(0,k)}\,\tilde\psi_\sigma(0,k)\,{\rm d}k=\eps^2\sum_\sigma\int_0^\infty\frac{\overline{\phi(\sigma k,k)}\,\psi(\sigma k,k)}{2k}\,{\rm d}k .
\end{equation}
$B$ extends continuously to $\Phi$. The physical Hilbert space is $\Hphy=\Hil_+\oplus\Hil_-$, with $\Hil_\pm\simeq L^2(\R_+,{\rm d}k)$, and $f_\sigma=Q_\sigma\psi$. On the positive-frequency sector, $f_+(k)\propto\sqrt k\,\tilde\chi(k)$, and $\|f_+\|^2$ is the Klein--Gordon norm of \cite{ACS}.
\end{proposition}

\begin{proof}
On $\D$ the function $G(y)=\sum_\sigma\int\overline{\tilde\phi_\sigma}\tilde\psi_\sigma\,{\rm d}k$ is smooth with compact support, so Fourier inversion gives $G(0)$. The last form follows from (\ref{eq:shellcoords}), or equivalently from $\delta(c/\eps^2)=\eps^2\sum_\sigma\delta(p-\sigma k)/(2k)$. Continuity follows from the trace bound. Let $\chi$ be the positive-frequency solution with Fourier data $\tilde\chi(k)=\psi(k,k)/(2k)$, the shell restriction produced by averaging. Then $f_+=\eps\sqrt{2k}\,\tilde\chi$ and $\|f_+\|^2=2\eps^2\int k|\tilde\chi|^2{\rm d}k$, the Klein--Gordon norm up to normalization.
\end{proof}

Group averaging thus reproduces the known sLQC physical inner product \cite{ACS,KLP}, now on an explicit test space with a continuous rigging map. The two frequency sectors are superselected: seeds supported in $p_\phi>0$ average into $\Hil_+$, which is the sector retained in sLQC. Proposition~\ref{prop:affine} carries over verbatim: $[\psi]\mapsto[Q\psi]$ is an affine bundle over $\PP(\Hil_+\oplus\Hil_-)$, modelled on $L^*\otimes\underline\N$ with $\N=\ker Q$ infinite-dimensional, and the fibres are again strictly larger than the orbits of $e^{-i\lambda\Chat}$. So does Proposition~\ref{prop:vertical}, with $e^{-t\eps^4y^2}$ in place of $e^{-t\eps^2y^2}$.

What does change is the behaviour of regularized averages. For a Gaussian regulator $e^{-\tau\lambda^2}$ the $\lambda$-integral produces, at each $k$, the convolution of $G(\cdot,k)$ with a kernel of width $\sqrt{2\tau}$, where $G$ is extended by zero below the edge. An $H^1$ function on $(-\kappa_k,\infty)$ satisfies $\sup_y|\tilde\psi_\sigma(y,k)|^2\le2\|\tilde\psi_\sigma(\cdot,k)\|^2_{H^1(-\kappa_k,\infty)}$, uniformly in $\kappa_k$, so dominated convergence gives the convergence of the regulated form to $B$ on all of $\Phi$. The convergence is not uniform in $k$, however. Near the edge the kernel loses a fraction of order $\frac12{\rm erfc}\bigl(\kappa_k/2\sqrt\tau\bigr)$ of the edge value of $G$, so the regulator must be narrow compared with the edge distance, $\sqrt\tau\ll\kappa_k$, and this fails as $k\to0$, where the shell approaches the edge.

\subsection{The canonical lift feels the edge}

The $\Phi$-orthogonal complement of $\N$ again defines a canonical section. Minimizing $\int_{-\kappa}^\infty(|g|^2+|g'|^2)\,{\rm d}y$ subject to $g(0)=1$ gives $g''=g$ on each side of the shell. With decay at $+\infty$ and the natural Neumann condition at the edge, the minimizer is
\begin{equation}\label{eq:gkappa}
g_\kappa(y)=\begin{cases}e^{-y}, & y\ge0,\\[2pt] \cosh(y+\kappa)/\cosh\kappa, & -\kappa\le y\le0,\end{cases}\qquad \|g_\kappa\|^2_{H^1}=1+\tanh\kappa .
\end{equation}

\begin{proposition}\label{prop:edge}
The canonical section is $(\sigma_*f)_\sigma(y,k)=g_{\kappa_k}(y)\,f_\sigma(k)$. With $n(\kappa)=\|g_\kappa\|^2_{L^2}$ and $D(\kappa)=2\cosh^2\kappa+2\kappa+\sinh2\kappa$,
\begin{equation}\label{eq:nkappa}
n(\kappa)=\frac{D(\kappa)}{4\cosh^2\kappa},\qquad \frac{\ev{\Chat}_{g_\kappa}}{\eps^2}=\frac{1-\kappa^2}{D(\kappa)},
\end{equation}
where the mean is taken in a single mode. Consequently:
\begin{enumerate}[(i)]
\item The canonical lift is not a product profile. It interpolates between a one-sided profile with $\|g\|^2_{H^1}\to1$ as $\kappa\to0$ and the clock-model profile $e^{-|y|}$, with norm $2$, as $\kappa\to\infty$.
\item The induced embedding $j_*:\Pphy\to\Ps$, $[f]\mapsto[\sigma_*f]$, is holomorphic but not K\"ahler isometric. The pullback of the kinematical Fubini--Study metric is the Fubini--Study metric of the Hermitian form $\langle f,n(\kappa_k)f\rangle$, which differs from $g_{\rm phy}$ unless $n(\kappa_k)$ is constant on the support of $f$. Since $n(\kappa)\in[\frac12,1.10]$, with $n\to\frac12$ as $\kappa\to0$ and $n\to1$ as $\kappa\to\infty$, the distortion is bounded but nonzero.
\item K\"ahler-isometric continuous sections exist, for instance with $g_\kappa$ replaced on $y\ge0$ by $e^{-a_\kappa y}$ and $a_\kappa$ chosen to keep $\|g\|_{L^2}$ constant, with $a_\kappa\in[\frac12,1.25]$. None of them is the canonical one. In sLQC, no section is both canonical for the test-space structure (\ref{eq:slqc-test}) and isometric for the quantum geometry; for the ideal clock, $e^{-|y|}$ was both.
\item Canonical seeds of modes near the edge, $\kappa<1$, have $\ev{\Chat}>0$: they lie on average on the side $p_\phi^2>\Theta$ of the shell. The bias vanishes at $\kappa=1$, is small and negative for $\kappa>1$, and tends to zero as $\kappa\to\infty$.
\end{enumerate}
\end{proposition}

\begin{proof}
The Euler--Lagrange equation and boundary conditions give (\ref{eq:gkappa}), and the $\Phi$-orthogonality of $\sigma_*\Hphy$ to $\N$ is the statement that $g_\kappa$ minimizes the norm at fixed shell value. Elementary integration gives $\int_{-\kappa}^\infty g_\kappa^2=D/(4\cosh^2\kappa)$ and $\int_{-\kappa}^\infty y\,g_\kappa^2=(1-\kappa^2)/(4\cosh^2\kappa)$. Item (ii) follows because a linear map $f\mapsto Mf$ pulls the Fubini--Study metric back to that of $\langle f,M^*Mf\rangle$. For (iii), the left-hand part of $g_\kappa$ carries $L^2$ mass $n(\kappa)-\frac12\le0.60$, so $a_\kappa=[2(\frac32-n(\kappa))]^{-1}\le1.25$ keeps the total mass equal to $1$ with uniformly bounded $H^1$ norm.
\end{proof}

\begin{figure}[t]
\centering
\includegraphics[width=\textwidth]{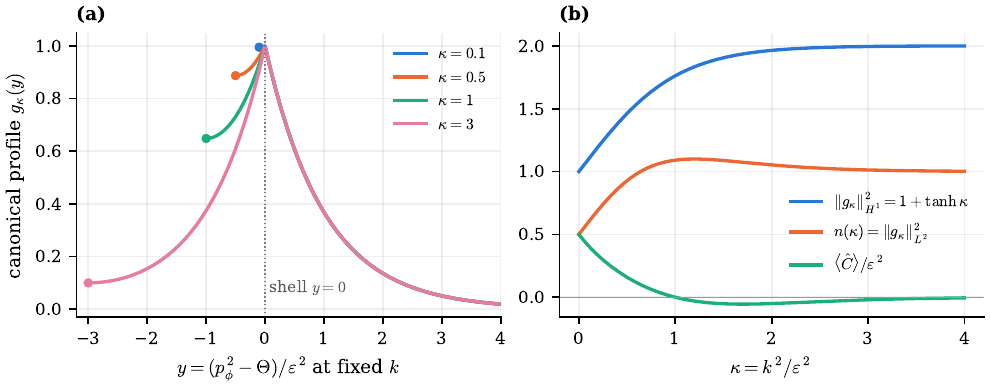}
\caption{Canonical lift in sLQC. (a) The minimal-norm profile $g_\kappa$ of (\ref{eq:gkappa}) for several values of the edge distance $\kappa=k^2/\eps^2$; dots mark the spectral edge $y=-\kappa$. (b) Its $H^1$ norm, its kinematical norm $n(\kappa)$ and the mean constraint of the canonical seed, (\ref{eq:nkappa}). For $\kappa\to\infty$ the clock-model values $(2,1,0)$ are recovered.}
\label{fig:slqc}
\end{figure}

The maximum of $n$ is $1.0998$, attained at $\kappa\simeq1.20$. Figure~\ref{fig:slqc} shows the profiles and the functions in (\ref{eq:nkappa}); all closed forms were checked numerically (see the data availability statement). Narrowed profiles still give Weyl sequences: for each $k$ one may compress $g_\kappa$ toward the shell inside $(-\kappa_k,\infty)$, so that $\Cq\to0$ while $Q\sigma f=f$. Approximate solutions therefore exist mode by mode, but for small universes they require profile widths below the edge distance $\kappa_k$.

Are these effects artifacts of the single scale $\eps$? One may object that a cleverer test norm, with a scale adapted to each mode, would remove them. The following proposition shows that it cannot remove both at once.

\begin{proposition}\label{prop:weights}
Consider the mode-wise weighted norms
\begin{equation}\label{eq:weighted}
\|\psi\|^2_{w,\ell}=\sum_\sigma\int_0^\infty{\rm d}k\;w_k\int_{-\kappa_k}^\infty\bigl(|\tilde\psi_\sigma|^2+\ell_k^2\,|\partial_y\tilde\psi_\sigma|^2\bigr){\rm d}y ,
\end{equation}
with $\ell_k$ bounded above and $w_k\ell_k$ bounded below, so that $Q$ remains continuous and the constraint group preserves the corresponding space. The rigging form $B$ is the same on $\D$. The canonical profile at mode $k$ is $g_{\kappa_k/\ell_k}(y/\ell_k)$, with kinematical norm $\ell_k\,n(\kappa_k/\ell_k)$ and mean constraint $\eps^2\ell_k(1-\kappa_k^2/\ell_k^2)/D(\kappa_k/\ell_k)$. The mean vanishes if and only if $\ell_k=\kappa_k$, and then the kinematical norm $\kappa_k\,n(1)$ varies with $k$. Hence no choice of weights makes the canonical section both K\"ahler isometric and centred on the shell at more than one value of $k$. For the ideal clock, the analogous canonical profile is $e^{-|y|/\ell_n}$, which is centred for every $\ell_n$ and isometric when $\ell_n$ is constant.
\end{proposition}

\begin{proof}
The half-line trace inequality gives $|f(0)|^2\le2\|f\|\,\|f'\|\le\ell^{-1}\bigl(\|f\|^2+\ell^2\|f'\|^2\bigr)$, so $Q$ is continuous when $w_k\ell_k$ is bounded below, and $\|\partial_y(e^{-i\lambda y}f)\|\le|\lambda|\|f\|+\|f'\|$ shows that the group preserves the space when $\ell_k$ is bounded above. The weight $w_k$ does not affect the minimizer. The substitution $y=\ell_ks$ turns $\int_{-\kappa}^\infty(|g|^2+\ell^2|g'|^2){\rm d}y$ into $\ell\int_{-\kappa/\ell}^\infty(|g|^2+|g_s|^2){\rm d}s$, whose minimizer at fixed $g(0)=1$ is $g_{\kappa/\ell}(s)$. Its $L^2$ norm in $y$ is $\ell\,n(\kappa/\ell)$ and its mean value of $y$ is $\ell$ times that of $g_{\kappa/\ell}$, which by (\ref{eq:nkappa}) vanishes only when $\kappa/\ell=1$. In the ideal-clock case the same substitution gives $e^{-|y|/\ell}$, which is even in $y$ and has $L^2$ norm $\ell$.
\end{proof}

Isometric choices do exist, since $\ell\,n(\kappa_k/\ell)={\rm const}$ can be solved for each $k$, but they are biased wherever $\ell_k\neq\kappa_k$. The incompatibility of the two properties is therefore a consequence of the edge, not of the choice of scale. What does depend on the scale is which modes count as `near the edge': with a single scale $\eps$, these are the small universes, $k\lesssim\eps$.

\subsection{Observables and the relativistic clock}

Bounded functions $A(k)$ of $\Theta$ act fibrewise in (\ref{eq:shellcoords}). They are kinematical Dirac observables that preserve $\Phi$ and intertwine $Q$. The momentum $\hat p_\phi$ also commutes with $\Chat$ and descends to multiplication by $\sigma k$, but in shell coordinates it acts as $\sigma\sqrt{\eps^2y+k^2}$, whose $y$-derivative diverges at the edge, so it does not preserve $\Phi$. Physical observables that mix $k$, such as the relational volume $\widehat V|_\phi$ of \cite{ACS}, act on $\Hphy$. Their kinematical extensions would have to interpolate across the edge, where the fibre spaces of $\Chat$ change, and we do not need them: every physical prediction depends on a seed only through $Q\psi$. For example, $\ev{\widehat V|_\phi}=V_+e^{\sqrt{12\pi G}\,\phi}+V_-e^{-\sqrt{12\pi G}\,\phi}$, with $V_\pm>0$ fixed by the physical state \cite{ACS}, is the same for every seed in the fibre.

Clock conditioning also changes. The clock is relativistic: $\hat p_\phi$ enters quadratically, so conditioning requires a choice of frequency sector \cite{HSL2}. In $\Hil_+$, the Page--Wootters state at clock reading $\phi$ is $e^{i\phi\sqrt\Theta}f_+$, which evolves unitarily in $\phi$. The slice $\chi(\cdot,\phi)$ of the Klein--Gordon solution corresponds instead to $(2\sqrt\Theta)^{-1/2}e^{i\phi\sqrt\Theta}f_+$, up to normalization conventions. Its kinematical norm does not depend on $\phi$, but it is not the physical norm: the map $f_+\mapsto\chi(\cdot,\phi)$ is not an isometry for the Klein--Gordon inner product. The factor $(2\sqrt\Theta)^{1/2}$ is the square root of the Jacobian $|\partial c/\partial p_\phi|=2k$ of the constraint at the shell, which appears in (\ref{eq:slqc-B}); for the ideal clock this Jacobian equals $1$. It is also the map between the group-averaging and Schr\"odinger pictures analysed in \cite{KLP}. sLQC therefore has four objects where the ideal clock had three:
\begin{itemize}[itemsep=0pt]
\item a normalizable seed $\sigma f\in\Hil$;
\item the generalized solution $\chi$;
\item its slice $\chi(\cdot,\phi)$;
\item the conditioned state $e^{i\phi\sqrt\Theta}f_+$.
\end{itemize}
Only the last carries the physical inner product at fixed clock reading, and none of them is a section of the averaging bundle.

\section{Discussion}\label{sec:discussion}

In this article we have studied the map that takes a normalizable kinematical seed to the physical state that group averaging produces from it. Let us summarize what we have learned, and then discuss what it implies, and what it does not imply, for gravitational systems.

\subsection{Two regimes, one picture}

When averaging is an orthogonal projection, the exact physical rays are at the same time an embedded K\"ahler submanifold and the base of a vector bundle of seeds whose zero section is fixed by the constraint solutions. The fibres are complex, they contain the gauge orbits and their complexifications, and they are in general much larger. The physical symplectic form pulls back to a form whose kernel is exactly the fibres, while the ambient form does not descend at all. A unitary flow descends to the base exactly when its generator commutes with the averaging projection, which, for bounded generators, is the Hessian criterion of the embedded description. The two descriptions of the physical state space, as a submanifold and as a base, are therefore compatible: they select the same observables, and they meet along the zero section.

For the clock model and for sLQC, physical histories are distributions, mean-ergodic averaging is zero, and the rigging construction on a test space gives an affine bundle. What is lost in passing from one regime to the other is precisely the zero section. Each section identifies the affine bundle with a vector bundle, and sections defined by a fixed seed profile embed the physical ray space into the kinematical one, isometrically in the clock model, but the embedded rays are seeds, not solutions. The test-space inner product singles out one such section, and narrowing the profile approaches the constraint shell through Weyl sequences. In sLQC the quadratic clock adds a spectral edge. The canonical section then depends on the mode, is not K\"ahler isometric, and biases the seeds of modes near the edge, while K\"ahler-isometric sections exist but are not canonical. In all regimes the physical metric and symplectic form are those of $\Pphy$. Pulling them back to representatives removes vertical directions without introducing a further quotient of physical states.

\subsection{What is canonical and what is not}

It is useful to collect the objects we have met according to the information they require. Table~\ref{tab:canonical} does so.

\begin{table}[ht]
\centering
\small
\renewcommand{\arraystretch}{1.2}
\begin{tabular}{p{0.37\textwidth}p{0.55\textwidth}}
\toprule
\textbf{Object} & \textbf{What it depends on} \\
\midrule
Physical rays $\Pphy$, their K\"ahler geometry, Dirac observables & The constraint and the physical inner product only \\
Averaging fibres & The averaging map; in the orthogonal case, only $\Hphy$ \\
Zero section & Exists only when physical states are normalizable \\
A section of the averaging bundle & A choice of seed family or profile \\
Canonical section & The test-space norm, including its scale $\eps$ \\
K\"ahler isometry of a section & The profile being mode-independent; fails for the canonical section in sLQC \\
Gradient flow of $\Cq$ & The operator chosen to represent the constraint \\
Clock-conditioned states & The choice of clock, and of frequency sector for a relativistic one \\
\bottomrule
\end{tabular}
\caption{Canonical structures versus structures that depend on additional choices.}
\label{tab:canonical}
\end{table}

Two entries deserve a comment. The gradient flow of $\Cq$ is defined by the constraint operator, not by the constraint surface: replacing $\Chat$ by $A\Chat$, with $A$ positive and commuting, leaves the physical states unchanged but changes the flow, and in the distributional case it also changes the rigging map. In sLQC, group averaging with the linear constraint $\hat p_\phi-\sqrt\Theta$ instead of $\hat p_\phi^2-\Theta$ produces an inner product without the factor $1/(2k)$ of (\ref{eq:slqc-B}), and the two are related precisely by $(2\sqrt\Theta)^{1/2}$, the map between the group-averaging and Schr\"odinger pictures analysed in \cite{KLP}. The Jacobian that separates the Page--Wootters state from the Klein--Gordon slice is thus the trace left by the quadratic representation of the constraint. The second comment concerns the canonical section. It is canonical only relative to the test space, and in sLQC it inherits the edge. Proposition~\ref{prop:weights} shows that the resulting tension between isometry and centring cannot be removed by reweighting, so it reflects the spectrum of the constraint and not the arbitrariness of the scale.

\subsection{The gradient flow, group averaging and Weyl sequences}

One of the lessons of \cite{Corichi2008} is that group averaging and the gradient flow of $\Cq$ are two faces of the same construction. In the bundle language this becomes very concrete. The flow is vertical: it moves seeds along their fibres and never changes the physical state they represent. In the orthogonal case it reaches the zero section, and group averaging is its endpoint. In the distributional case there is no endpoint in $\Ps$. The flow narrows the seed profile toward the shell, produces Weyl sequences of approximate solutions, and accumulates on Dirac states that are not vector states at all. The Gaussian-regulated average (\ref{eq:regulated}) is the overlap along this flow, and the physical inner product is the coefficient of its decay. From this perspective, the different procedures that one finds in the literature for constructing physical states, namely Haar averaging, mean-ergodic averaging, refined algebraic quantization, Klauder's projections onto small values of the constraint, and the master constraint, are different ways of moving along, or jumping across, the same fibres.

\subsection{Relativistic clocks and the problem of time}

The comparison between the ideal and the relativistic clock is perhaps the most physically relevant outcome of this work. With an ideal clock, three objects must be kept apart: a clock-conditioned state, a normalizable seed, and a generalized solution. With a relativistic clock there are four, because the slice of the solution and the conditioned state no longer coincide; they differ by the Jacobian $(2\sqrt\Theta)^{1/2}$. Any geometric statement about `the state of the universe at a given value of the scalar field' must therefore specify which object is meant. This is consistent with the analyses of \cite{KLP,HSL1,HSL2}, which compare physical inner products and relational descriptions of evolution. The present analysis complements those studies by isolating the geometry of the freedom in the seeds, and by showing where that freedom interacts with the choice of clock.

\subsection{Complexified gauge orbits and the classical gauge directions}\label{sec:complexified}

There is a further aspect of the averaging fibres that deserves attention. We have seen that, within the coherent family of Section~\ref{sec:oscillators}, each averaging fibre is exactly one complexified gauge orbit. Its real part is the common-phase circle of the coherent label. On the classical constraint surface $R=N$ this circle is the Hopf fibre, which is the classical gauge orbit, i.e.\ the null direction of the symplectic form restricted to the constraint surface. Its imaginary part is the radial direction, transverse to the constraint surface. In a K\"ahler manifold these two directions are related by the complex structure, since the gradient of a function is $J$ times its Hamiltonian vector field. Group averaging discards both: it maps a whole complexified orbit, and much more, to a single physical ray.

The same pair of directions appears in a different guise in \cite{CorichiGauge}, where the question is how the unreduced classical geometry is recovered from a positive master constraint. There it is shown that the linearization $L$ of the master Hamiltonian vector field at the classical constraint surface maps transverse displacements into gauge directions, and that the image of $L$ is the characteristic distribution of the surface. For a quadratic master constraint in a K\"ahler frame, $L$ maps the imaginary gauge directions $JX_{C_i}$ onto the real ones $X_{C_i}$. Thus the two constructions see the same structure from opposite sides. Group averaging \emph{identifies} the real and imaginary gauge directions, together with the rest of the fibre, while the master flow \emph{relates} them, converting the one into the other, and it is in the response of kinematical seeds near the constraint surface that the classical gauge geometry becomes visible. In both cases the relevant states are kinematical seeds, not physical solutions; the physical ray space itself has no degenerate directions, and it is only before the quotient is taken that the gauge geometry can be seen.

\subsection{Scope and outlook}

The scope of each statement should be kept explicit. Proposition~\ref{prop:bundle} requires a nonzero closed physical subspace and an orthogonal projection, realized by compact groups, by mean-ergodic averages when $0$ is an eigenvalue, by F\o lner averages of amenable groups, or by heat-kernel averaging. It includes several constraints and non-Abelian groups whenever their invariant subspace is the physical one. The clock and sLQC results use explicit, surjective trace maps on the Sobolev spaces (\ref{eq:testspace}) and (\ref{eq:slqc-test}). The canonical sections depend on the scale $\eps$ of these norms, while all physical predictions do not. For a general positive rigging form, the quotient $\Phi/\N$ may require completion, and a continuous linear splitting need not exist \cite{GiuliniMarolf1,GiuliniMarolf2}. Affine fibres exist algebraically over the image rays, but local triviality and coverage of the completed physical ray space require additional hypotheses, which the clock model shows can be verified.

Both the ideal clock and sLQC have a global deparametrization and a single self-adjoint constraint generating $\R$, and sLQC is homogeneous and exactly solvable. A gravitational system need not admit a global clock, or a Lie-group action implementing its full constraint algebra, and we do not infer an averaging bundle for full general relativity from these models. Where group averaging is not available, the gradient flow of the master function \cite{Corichi2008} still is, and since it is vertical in every case we have examined, it is a natural candidate for exploring the seed-to-state map when the constraints have structure functions. The spectral-edge effects of Section~\ref{sec:slqc}, on the other hand, rely only on the clock momentum entering quadratically, and should persist in other deparametrized models with relativistic clocks, such as cosmologies with spatial curvature or a cosmological constant. It would be interesting to see whether the edge leaves an imprint on the construction of physical semiclassical states, along the lines of \cite{ABC}, for universes that are small on the scale of the test space.

What the examples establish is which parts of the relation between embedded solutions, averaging fibres and lifts are canonical, and which depend on the seed family, on the test-space structure or on the clock. Needless to say, these matters need to be further studied, in particular in models with a richer constraint algebra. We hope that the geometric viewpoint put forward here, besides clarifying the standard constructions, might help in disentangling the choices that enter the comparison of physical inner products and of descriptions of evolution in quantum gravity.

\section*{Acknowledgments}
I thank Steph and Fide for comments and support. This work received no external funding.

\section*{Data availability}
No new data were created or analysed in this study. Scripts verifying the closed forms of Sections~\ref{sec:clock} and \ref{sec:slqc} and generating Figure~\ref{fig:slqc} are provided as ancillary files.

\end{document}